# Nondestructive testing of grating imperfections using grating-based X-ray phase-contrast imaging

Shenghao Wang[1, #] and Can Zhang[2]

[1]*Key Laboratory of Materials for High Power Laser, Shanghai Institute of Optics and Fine Mechanics, Chinese Academy of Sciences, Shanghai, China, 201800*
[2]*National Synchrotron Radiation Laboratory, University of Science and Technology of China, Hefei 230027, China*

[#] Corresponding author: wangshenghao@siom.ac.cn

We reported the usage of grating-based X-ray phase-contrast imaging in nondestructive testing of grating imperfections. It was found that electroplating flaws could be easily detected by conventional absorption signal, and in particular, we observed that the grating defects resulting from uneven ultraviolet exposure could be clearly discriminated with phase-contrast signal. The experimental results demonstrate that grating-based X-ray phase-contrast imaging, with a conventional low-brilliance X-ray source, a large field of view and a reasonable compact setup, which simultaneously yields phase- and attenuation-contrast signal of the sample, can be ready-to-use in fast nondestructive testing of various imperfections in gratings and other similar photoetching products.

## 1. INTRODUCTION

X-ray phase-contrast imaging, which uses phase shift as the imaging signal, can provide remarkably improved contrast over conventional absorption-based imaging for weakly absorbing samples, such as biological soft tissues and fibre composites.[1-4] Over the last 20 years, several X-ray phase-contrast imaging methods has been put forward, they can be classified into crystal interferometer,[5] free-space propagation,[6, 7] diffraction enhanced imaging,[8, 9] and grating interferometers.[10, 11] Although many excellent experiments research were accomplished based on these techniques,[12-19] none of them has so far found wide applications in medical or industrial areas, where typically the use of a laboratory X-ray source and a large field of view are required. In 2006, Pfeiffer et al.[20] first developed and demonstrated a Talbot-Lau interferometer in the hard X-ray region with a low-brilliance X-ray source, this can be considered as a great breakthrough in X-ray phase-contrast imaging, because it showed that phase-contrast X-ray imaging can be successfully and efficiently conducted with a conventional, low-brilliance X-ray source, thus overcoming the problems that impaired a wider use of phase-contrast in X-ray radiography and tomography, and many potential applications in biomedical imaging of this technique have been studied.[21-28]

However, little attention has been paid to the potential application of Talbot-Lau interferometer in industrial nondestructive testing area[29, 30], where conventional absorption-based X-ray imaging technique is not competent in some cases. The aim of this work is to demonstrate how we use both the phase- and absorption-contrast imaging signal, simultaneously generated by a grating-base X-ray phase-contrast imaging setup, to nondestructively detect the imperfections of X-ray grating, such as the flaws resulting from uneven ultraviolet exposure phenomenon of the photoetching machine, and the defects in electroplating process.

## 2. MATERIALS AND METHODS

2.1 Experimental setup and working principle

The grating-based X-ray phase-contrast imaging experiments were carried out at the National Synchrotron Radiation Laboratory (NSRL) of the University of Science and Technology of China (USTC), in Hefei, China. Fig. 1(a) is the mechanical structure of the grating-based X-ray phase-contrast imaging setup. It is mainly made up of an X-ray tube, an X-ray flat panel detector and three micro-structured gratings, which are assembled on multi-dimensional motorized optical displacement tables assembled by 21 motorized positioning stages (15 translation stages, 3 rotary stages and 3 goniometric stages) (Beijing Optical Century Instrument Co., Ltd, China). Because of the demanding precision, an ultra-precision piezoelectric translation stage (Micronix Inc., California, America.) with an encoder has been used for the phase-stepping scan.

The X-ray tube is a cone beam X-ray source (YXLON international GmbH, Hamburg, Germany), and we use the round, stable focal spot (1.0 mm in diameter) on a tungsten target anode. The usable voltage is 7.5-160 kV, and the X-ray tube is cooled using a commercial available centrifugal chiller (HTCY Technology, Beijing, China). The source grating G0 (period 100 μm, gold height 50 μm, size 1×1 cm$^2$) is positioned about 10 mm from the emission point inside the X-ray source, the beam splitter grating G1 (period 50 μm, gold height 50 μm, size 10×10 cm$^2$) is placed 270 mm behind the gantry axis, and the analyzer grating G2 (period 100 μm, gold height 50 μm, size 10×10 cm$^2$) is positioned in contact with the flat-panel detector, the distance between G1 and G2 is 270 mm. All the three gratings were produced by the LIGA process, involving EUV photoetching and electroplating. The X-ray images were captured using a flat penal detector (PerkinElmer Inc., Waltham, Massachusetts, USA) with an effective receiving area of 20.48×20.48 cm$^2$ and 0.2×0.2 mm$^2$ pixel size (without binning). The field of view (FOV) is limited by the size of G2 (10×10 cm$^2$), considering the geometric magnification, the FOV at the sample (located near G1 grating) is 5×5 cm$^2$, and the effective pixel size of the sample is 0.1×0.1 mm$^2$.

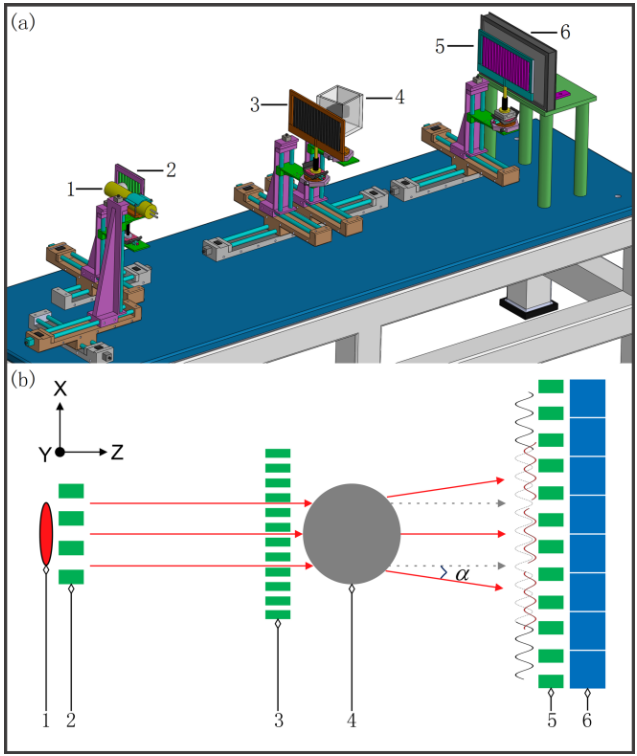

FIG. 1. (Color online). (a) Mechanical structure and (b) working principle of the grating-based X-ray phase-contrast imaging setup at NSRL-USTC. 1. X-ray source, 2. Source grating (G0), 3. Beam splitter grating (G1), 4. Sample, 5.Analyzer grating (G2), 6. X-ray flat panel detector.

As illustrated in Fig. 1(b), working principle of the imaging system is briefly described as follow, the source grating G0, an absorbing mask with transmitting slits, placed close to the X-ray tube anode, creates an array of line sources. Grating G1 acts as a beam splitter, through the direct geometric projection, shadow of G1 forms in the plane of grating G2. The differential phase-contrast image information process achieved by the two gratings G1 and G2, essentially relies on the fact that sample placed in the X-ray beam path causes slight refraction of the beam transmitted through the object. The fundamental idea of differential phase-contrast imaging depends on locally detecting these angular deviation, the angular $\alpha$ is proportional to the local gradient of the object's phase shift, and can be quantified as:

$$\alpha = \frac{\lambda}{2\pi} \frac{\partial \Phi(x,y)}{\partial x}. \qquad (1)$$

Where $\Phi(x,y)$ is the phase shift of the wave front, and $\lambda$ represents wavelength of the radiation. Determination of the refraction angle can be achieved by phase-stepping (PS),[31] a typical measurement strategy, which contains a set of images taken at different positions of the grating G2. When G2 is scanned along the transverse direction, the intensity signal in each pixel in the detector plane oscillates as a function of the grating position. By Fourier analysis, for each pixel, the shift curve (SC) of these oscillations, sample's conventional transmission and refraction signal can be simultaneously retrieved. Much detailed description of the imaging system's theoretical basis can be found here.[32]

### 2.2 Image acquisition and data post-processing

The experiments were performed without an extra sample in the beam path, and with 40 kV X-ray tube acceleration voltage and a current of 22.5 mA. After fine alignments of the three gratings, infinite and even moire fringes were generated, then 21 steps were adopted during the PS scan, and for each step, 10 raw images were captured to reduce statistical and systemic noise, note that exposure time of a single image is 2 seconds, this resulted in a total exposure time of about 10 minutes for one PS scan.

Based on the acquired dataset, transmission $\mu$ and refractive angle $\theta$ signal of the sample can be simultaneously retrieved by:

$$\mu(m,n) = \frac{\sum_{k=1}^{N} I_k(m,n)}{N}, \qquad (2)$$

$$\alpha(m,n) = \frac{p_2}{2\pi d} \cdot \arctan\left[\frac{\sum_{k=1}^{N} I_k(m,n) \cdot \sin\left(2\pi \frac{k}{N}\right)}{\sum_{k=1}^{N} I_k(m,n) \cdot \cos\left(2\pi \frac{k}{N}\right)}\right]. \qquad (3)$$

Here $N$ is the steps of PS scan in one period of grating G2, $I_k(m,n)$ is gray value of pixel $(m,n)$ at the $k^{th}$ step, $p_2$ is the period of grating G2, and $d$ represents the distance between grating G1 and G2.

Data post-processing and signals retrieval were accomplished by a LabVIEW-based software platform,[33] where unified management and control of all the motorized positioning stages are also achieved, moreover, automatic image acquisition during PS scan and other custom features are available.

### 3. RESULTS AND DISCUSSIONS

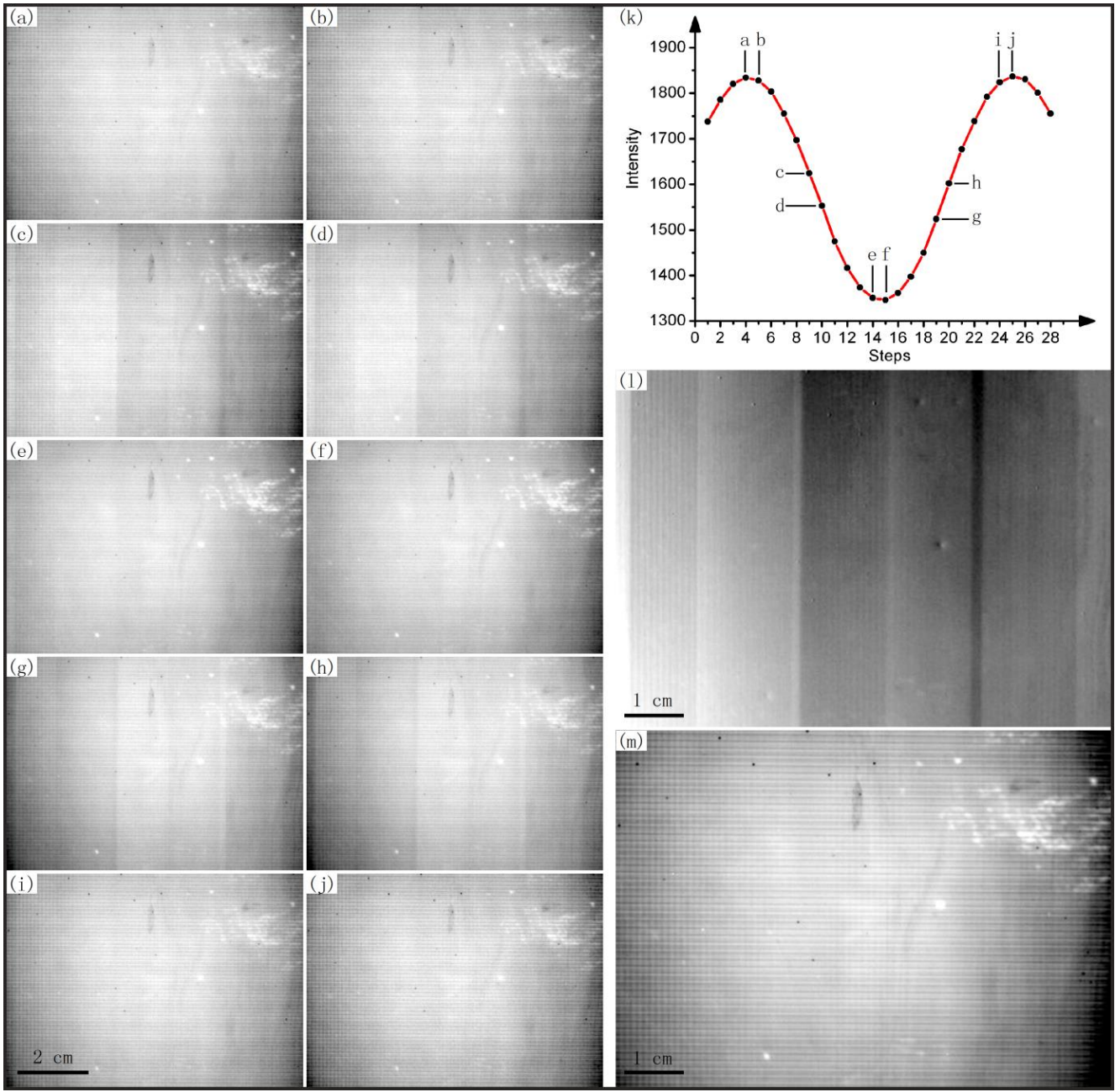

FIG. 2. X-ray imaging results of the system's background and the raw data. Top right (k) is the sinusoidal SC (generated by the mean gray value of all the pixels in the active FOV) of the PS scan, a-j represent some marked positions in the SC (a, b: left peak; c, d: left waist; e, f: wave trough; g, h: right waist; i, j: right peak.), and raw image of PS scan at each selected position is shown in (a)-(j), respectively. (l) is the retrieved X-ray differential phase-contrast image and (m) is conventional absorption-based signal. All the images are displayed on a linear gray scale and are windowed for optimized appearance.

Fig. 2 is the X-ray imaging results of the system's background and part of the raw images in PS scan. Fig. 2(k) illustrates SC of the PS scan, a-j represent some marked positions in the SC (a, b: left peak; c, d: left waist; e, f: wave trough; g, h: right waist; i, j: right peak.), and raw projection image at each selected position is shown in Fig. 2(a)-Fig. 2(j), respectively. The retrieved attenuation-based image is shown in Fig. 2(l), while Fig. 2(m) depicts the differential phase-contrast image signal. It is found that clear vertical stripes with large period (about 2cm) exist in Fig. 2(l), the phase-contrast imaging, while from Fig. 2(m), the corresponding absorption signal, we can hardly detect their existence. At the same time, we observe the location of white dispersive spots in the upper right part of the view from the conventional absorption-based image, and on the contrary, they are completely invisible in the related phase-contrast signal.

The vertical strips with large period is a distinct and interesting discovery by the retrieved X-ray phase-contrast image of the system's background, it can be explained by something unusual appear in the beam path. Considering that the experiment was performed without an extra sample in the beam path, the only possibility is that the "something unusual" are the photoresists and silicon wafers supporting the Au grids of the three X-ray transmission gratings in the imaging system. Fig. 3 illustrated TEM of cross section through grating G2, we can see that the Au grids (in bright white) are supported by the photoresist (in gray) and the silicon wafer (in light white), while the Au grids act as an important medium in phase-contrast signal retrieval, both the still existing photoresists and the silicon wafer can be regarded as the sample under analysis in the above experiment.

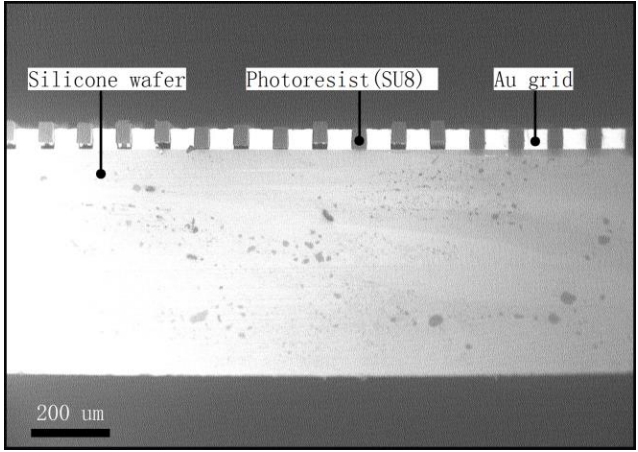

FIG. 3. TEM of cross section through the analyzer grating G2. Some photoresists grids has jumped out of the silicon wafer, this is because of the mechanical shock when we cut the wafer for the TEM test.

In order to judge which grating contributed the existence of the observed vertical strips and the white dispersive spots, we arranged another two groups of experiments after fine alignment of the three gratings. Group (1): Move grating G1 in X direction (as shown in Fig. 1) to the position X=0/-1/-2/-3 mm, respectively, and keep other optical components still, at each position of G1, conduct the PS scan and retrieve the signals. Group (2): Move grating G2 in X direction to the position X=0/2/4/6 mm, respectively, other elements hold still, and at each position of G2, perform the PS scan and calculate the absorption- and phase-contrast imaging signals.

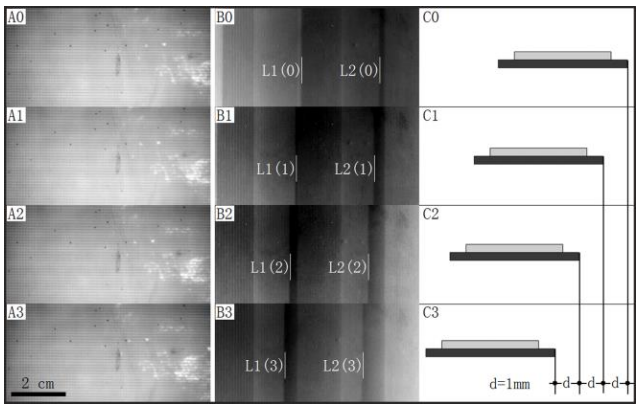

FIG. 4. Experimental results of group (1). C0, C1, C2, C3 depicts the relative displacement of G1 at the four positions, (other optical elements in the beam path are kept still.) A0, A1, A2, A3 are the corresponding retrieved absorption-based signal of each position, while B0, B1, B2, B3 are the differential phase-contrast imaging. All the images are displayed on a linear gray scale and are windowed for optimized appearance.

Fig. 4 is the experimental results of group (1). Fig. 4(C0), Fig. 4(C1), Fig. 4(C2), and Fig. 4(C3) depicts the relative displacement of G1 at the four different positions. Fig. 4(A0), Fig. 4(A1), Fig. 4(A2) and Fig. 4(A3) are the corresponding retrieved absorption-based images of each position, while Fig. 4(B0), Fig. 4(B1), Fig. 4(B2) and Fig. 4(B3) are the differential phase-contrast signals. From all the four phase-contrast images, we can see clear vertical strips, and some of them stay still, while the others have a horizontal shift (marked as L1 and L2), considering that grating G1 has a relative shift in X direction at each step. Judgment can be reached that L1 and L2 have the same moving direction with grating G1.

TABLE I. Quantitative analysis of L1 and L2 of group (1).

| Position of G1 | C0 | C1 | C2 | C3 |
|---|---|---|---|---|
| Relative shift of G1(mm) | 0 | -1 | -2 | -3 |
| X position of L1(L2) (1 pixel=0.2mm) | 545 (689) | 535 (679) | 524 (668) | 514 (658) |
| Calculated relative offset of L1(L2) | 0 (0) | -2.0 (-2.2) | -4.2 (-4.2) | -6.2 (-6.2) |

A quantitative analysis of the marked strips' shift are available at TABLE I. If we take into consideration that G1 has a little higher than 2× magnification in the plane of the detector (The distance between G1 and G2 is equal to that of G1 and G0, because of the non-negligible distance from G2 to the detector, the geometric magnification is a little larger than 2×), it would be found that L1 and L2 have almost the same motion with G1 at different steps. Thus we can say that L1 and L2 are contributed by G1. Meanwhile, if we have an eye on the white dispersive spots on right of the four absorption-based images, also we can deduce that these white dispersive spots are located at grating G1 as they have the same moving direction with G1 (quantitative analysis is not provided here).

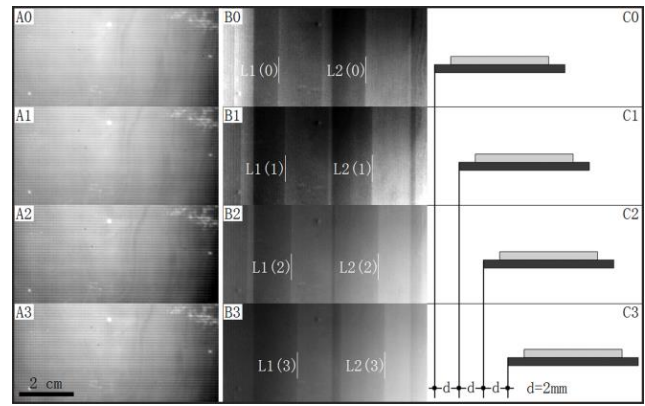

FIG. 5. Experimental results of group (2). C0, C1, C2, C3 depicts the relative displacement of G2 at the four positions, (other optical elements in the beam path are kept still.) A0, A1, A2, A3 are the corresponding retrieved absorption-based signals of each position, while B0, B1, B2, B3 are the differential phase-contrast image. All the images are displayed on a linear gray scale and are windowed for optimized appearance.

Fig. 5 is the experimental results of group (2), and TABLE II shows the quantitative analysis of the marked strips' horizontal offset. With a similar analyze, judgment can be made that the marked two vertical strips are located on grating G2. And we can also confirm the deduction in group (1) that the white dispersive spots are contributed by G1 because they share the same position with G2's motion, as shown in Fig. 5(A0), Fig. 5(A1), Fig. 5(A2), and Fig. 5(A3).

TABLE II. Quantitative analysis of L1 and L2 of group (2).

| Position of G2 | C0 | C1 | C2 | C3 |
|---|---|---|---|---|
| Relative shift of G2(mm) | 0 | 2 | 4 | 6 |
| X position of L1(L2) (1 pixel=0.2mm) | 488 (648) | 498 (658) | 510 (669) | 520 (680) |
| Calculated offset of L1 (L2) | 0 (0) | 2.0 (2.0) | 4.4 (4.2) | 6.4 (6.4) |

We can make the judgment, based on the above two groups of experiments, that the observed vertical strips with large period in the phase-contrast images are contributed by both grating G1 and G2, we think the strips stem from the uneven ultraviolet exposure phenomenon of the photoetching machine, G0 does not have a contribution to the strips and corresponding tests were not performed, because we think the uneven exposure phenomenon is negligible when fabricating G0 (the size of G0 is 400 times smaller than that of G1 and G2). In ultraviolet exposure process of grating fabrication, if a very tiny angle exists between the mask and the silicon wafer, equal thickness interference would happen, resulting in periodical varying exposure intensity in the plane of the photoresist, and thus impose a negative effect on the perfections of exposure and of the final Au grids structures. We think this finding is useful for evaluating the exposure uniformity of the photoetching machine. Meanwhile, it was found and proved that the observed white dispersive spots in the FOV are located in grating G1, and they can be regarded as electroplating flaws.

## 4. DISCUSSION

The experimental results demonstrate, compared with conventional absorption-based X-ray imaging, the remarkably enhanced X-ray phase-contrast imaging's performance for investigating the tiny discrimination of photoresist (SU8), and this phenomenon can be explained by that, low molecular weight SU8 consists of a polymeric epoxy resin by dissolving in an organic solvent and adding a photo acid generator,[34] and the sensitivity of phase-contrast imaging is about 1000 times higher than that of the conventional absorption-based signal for materials made up of low-Z elements in hard X-ray region.[2] Also the research shows the absorption-based imaging's strong discriminability for resolving electroplating imperfections. By combining both the attenuation- and phase-contrast signal, simultaneously yielded by a grating-based X-ray phase-contrast imaging setup, significantly more and unique information than any of the techniques alone would be provided.

We think potential application of grating-based X-ray phase-contrast imaging technique in nondestructive testing of grating imperfections and other similar photoetching products can be ready-to-use in the following described manner. Place the investigated products close to grating G1 in the beam path, and perform PS scan, then pure phase-contrast signal of the sample can be obtained, after the retrieved phase-contrast signal (with sample) subtract the signal (without sample), and once the absorption signal (with sample) divide the signal (without sample), pure absorption-based image of the inspected sample can be available. If we previously acquire and store the phase- and absorption-contrast image (without sample), both signals of the inspected sample would be available only after a PS scan.

In the previous experiments, 21 steps were adopted during the PS scan, and for each step, 10 raw images were captured, this resulted in a total exposure time of about 10 minutes. We want to point out that the exposure time can be greatly reduced by: (1) Fabricating gratings with higher Au grids, this would increase the fringe visibility and thus improve the sensitive of signal retrieval, 5-8 steps in PS scan would be adequate for successful and effective information retrieval once a >60% fringe visibility is reached (fringe visibility of the present confirming is only 15%); (2) Using rotating anode X-ray generators with a power of a few kW, the exposure time for one image and the number of images to average at one step of PS scan can be remarkably reduced. With these two improvements, a total exposure time of about 10 seconds for the testing one sample can be anticipated.

Considering that speed is a very important factor in industrial areas, one potential method based on this technique to enable fast nondestructive testing is a single shot strategy, in which a single X-ray projection image is captured on condition that grating G1 and G2 are aligned such that the SC walks around the left/right waist position. As shown in Fig. 2, Fig. 2(c) and Fig. 2(d) represent the raw projection images when SC walks around the left waist position, while Fig. 2(g) and Fig. 2(h) are those of the right waist. From each image, clear vertical strips and electroplating defects can be detected, and they have a similar visual appearance as these in the finally retrieved absorption- and phase-contrast signals. This can be explained by that absorption and refraction signals are mixed and best visual effect are delivered at the waist position of SC, some theoretical analysis of this phenomenon can be referred here.[35] Important benefits arise from this single shot strategy is that it eliminates the need for PS scan. Therefore, a greater reduction (about 80% if we adopt 5 steps PS scan) in the total exposure time for inspecting one sample can be promising, and a greater mechanical stability of the system can be obtained, moreover, cost can be remarkably reduced because the motor for high precision PS scan usually is very expansive, compared with the other mechanical components.

## 5. CONCLUSION

In conclusion, we have shown that grating-based X-ray phase-contrast imaging technique, which simultaneously yields phase- and absorption-contrast imaging signal, can be successfully used to effectively resolve the uneven photoresists, and other imperfections in X-ray transmission grating. The results demonstrate that grating-based X-ray phase-contrast technique, with a low-brilliance tube-based X-ray source, a large field of view and a reasonable compact setup, can be ready-to-use in fast nondestructive testing of grating quality and other similar photoetching products.

**ACKNOWLEDGEMENT**

The authors would thank WenXing Chen for the TEM tests, and Ying Xiong, Jian Chu for many fruitful discussions. This work was partly supported by the Major State Basic Research Development Program (2012CB825800), the Science Fund for Creative Research Groups (11321503), the Knowledge Innovation Program of The Chinese Academy of Sciences (KJCX2-YW-N42), the National Natural Science Foundation of China (61705246) and the Fundamental Research Funds for the Central Universities (WK2310000021).